\documentclass[aps,prapplied,reprint,groupedaddress,floatfix]{revtex4-2}

\usepackage{graphicx}
\graphicspath{ {images/}} 
\usepackage{dcolumn}
\usepackage{bm}
\usepackage[caption=false]{subfig}
\usepackage{amsmath}
\usepackage{amsfonts}
\usepackage{footmisc}

\begin{document}

\title{Spin-Type Photonic Topological Insulators on a Rhombic Lattice}

\author{Robert J. Davis}
\email{rjdavis@ucsd.edu}
\author{Daniel F. Sievenpiper}
\email{dsievenpiper@ucsd.edu}
\affiliation{Electrical and Computer Engineering Department, University of California San Diego, La Jolla, California 92093, USA}

\date{\today}

\begin{abstract}
A simplified model of a metallic spin-type photonic topological insulator on a rhombic lattice is presented and analyzed. Instead of the more commonly used hexagonal unit cells, a reduced symmetry rhombus is employed, which is both simpler and allows for easier integration into more traditional microwave systems. The non-trivial nature of the transport is shown via direct calculation of the structure's spin-projected Berry curvature and Wilson loop spectra, as well as by a systematic investigation of a reduced symmetry tight-binding model based off the Kane-Mele Hamiltonian. Device implementations are shown for a range of non-trivial configurations. 
\end{abstract}

\maketitle


\section{Introduction}
Symmetry breaking has wide sweeping consequences across physics, and is the foundation underlying the modern theories of topological classification of matter. For the original studies involving the quantum Hall effect\cite{kosterlitz_ordering_1973,thouless_quantized_1982}, the crucial symmetry  was time reversal symmetry ($\mathcal{T}$ symmetry), and for the later discoveries of the quantum spin Hall effect \cite{kane_quantum_2005,kane_z_2005} inversion symmetry ($\mathcal{I}$) was included, along with the effects of spin. It was later shown that crystallographic symmetries can also have a profound effect on topological phenomena in condensed matter systems \cite{fu_topological_2011,ando_topological_2015}. The ready application of these symmetries onto the experimentally simpler bosonic platforms (photonics \cite{haldane_possible_2008, wang_reflection-free_2008} and phononics \cite{liu_topological_2020} in particular) have led to wide sweeping discoveries with myriad practical applications, including nonreciprocal waveguides for $\mathcal{T}$-breaking systems \cite{wang_observation_2009}, spin-filtered routers \cite{cheng_robust_2016}, and robust cavity modes \cite{ota_photonic_2019}. 

One electromagnetic platform in particular \cite{khanikaev_photonic_2013}, which employs the additional symmetry of duality of the electric and magnetic fields to generate pseudospin pairs, is especially useful for introducing topological behaviors inside practical waveguiding systems. In these so-called photonic topological insulators, the added degree of freedom of the hybridized electric and magnetic profiles allows for a $\mathcal{T}$-invariant platform that can be directly mapped onto the original Kane-Mele Hamiltonian of the electronic quantum spin Hall effect. In the original proposal the electromagnetic (EM) duality condition was achieved by a quasi-2D system of coupled split-ring resonators placed into a hexagonal lattice. The $C_{6v}$ symmetry of the system leads to a symmetry protected 2-fold degeneracy at the $K/K'$ points of the Brillouin zone, which is later broken by an introduction of the bianisotropic coupling of the $E$ and $H$ fields. It was also later shown that this effect could be replicated in parallel-plate waveguides \cite{ma_guiding_2015} and radiative dielectric structures \cite{slobozhanyuk_near-field_2019}.

It was later shown \cite{bisharat_electromagnetic-dual_2019} that the EM duality condition can be engineered for planar surfaces via Babinet's principle  \cite{born_principles_1999}. In that work, the topologically nontrivial mode is confined to a quasi-1D "line wave" on a metal-on-dielectric platform, thereby allowing for easier fabrication in the lab with traditional PCB techniques. The underlying physics, and the needed mapping to the Kane-Mele Hamiltonian, is preserved in the system, as it maintains the same set of effective symmetries ($\mathcal{T}$ and $C_{6v}$ most critically). 

In this paper, we show how the reduction of the point group symmetry from $C_{6v}$ (hexagons) down to $C_{2h}$ (rhombi) maintains the same robust edge states, while possessing a number of unexpected properties with regards to its topological features. As such a unit cell can define edges that are straight lines, it also provides a considerably easier to use system for applications in waveguiding. This work builds upon our previously published work on this unit cell \cite{davis_robust_2021}. 

The paper is organized as follows: In Section \ref{sec:top} we introduce the model and analyze its topological features in reciprocal space numerically. We show how the reduction in point group symmetry has consequences on the Berry curvature and Wilson loop spectra. Then in Section \ref{sec:ham} we derive an effective Hamiltonian of our model and show how it relates to the Kane-Mele Hamiltonian. In Section \ref{sec:angle}, we show how altering the angle of the rhombus (thereby reducing the symmetry further still) changes the band structure as well as the topological behavior, eventually breaking down at the limits. In Section \ref{sec:dev} we showcase a range of finite lattices that show how the topological features discussed can be exploited into practical devices.  Finally, we provide some conclusions and suggestions for future work.

\begin{figure}
	\centering
	\includegraphics[width=\linewidth]{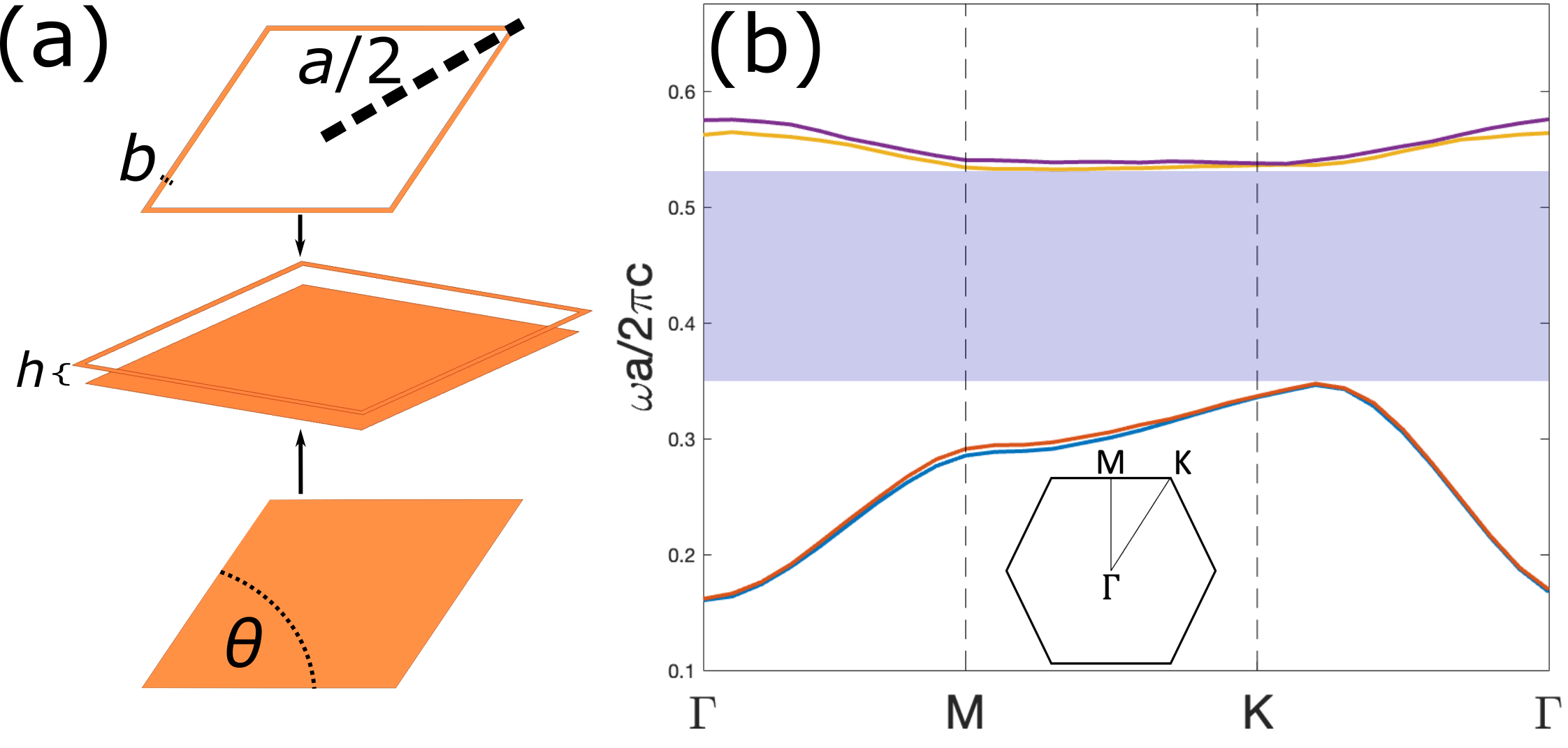}
	\caption{Duality spin rhombic unit cell and related band structures. (a) The unit cell composition, with period $a=20$ mm, thickness $h = 1.57$ mm, and cell boarder width $b = 0.43$ mm. Top shows the metallic patches, bottom shows the metallic frames, and the middle shows the full structure. (b) The photonic band structure of the full duality cell seen in (a). Inset is the Brillouin zone used, and the shaded region shows the complete bandgap. }
	\label{fig:rhomCell}
\end{figure}

\section{Electromagnetic Model and Topological Behavior}\label{sec:top}

For the original studies of spin-type photonic topological insulators (hereafter called spin PTIs), the basic models assume a direct mapping to the Kane-Mele Hamiltonian \cite{kane_quantum_2005} as their basis, and consist of a honeycomb lattice of cells that possess an internal degree of freedom owing to polarization, which can then be altered by bianisotropy \cite{khanikaev_photonic_2013}. This mapping retains as much symmetry as allowable, namely the crystallographic symmetry and time reversal symmetry for spin 1/2 particles. The former is achieved by geometry (e.g., selecting a hexagonal unit cell obeying $C_{6v}$ point group symmetry), while the later is enabled by the hybridized electromagnetic polarizations. For this study, we retain the electromagnetic duality (and therefore the pseudospin degree of freedom), while reducing the point group symmetry. 

Inspired by Ref. \cite{bisharat_electromagnetic-dual_2019}, the unit cell used is shown in Fig. \ref{fig:rhomCell}(a), with corresponding band structure shown in Fig. \ref{fig:rhomCell}(b). It consists of two layers of periodically arranged metal, one layer being connected metal frames, the other disconnected metal patches. The frame surface supports transverse magnetic surface modes, while the patches support transverse electric modes. As each is the geometric dual of the other, these act equally but oppositely on the electric and magnetic fields \cite{van_kruining_conditions_2016}, and as such result in equal mode dispersions. When the separation between the two surfaces is made small ($h<<a$), there is an effective bianisotropic coupling $\chi$ that hybridizes the modes. 

In contrast to the original design of Ref. \cite{bisharat_electromagnetic-dual_2019}, the point group of the unit cell (here $C_{2h}$ when considering the dual modes) does not admit 2D irreducible representation, and therefore will not have any enforced Dirac cone at $K$. A complete bandgap is therefore formed for both the TE and TM surface mode structures in the absence of bianisotropic coupling. Nevertheless, the EM duality condition of $\mu=\epsilon$ is still satisfied via the Babinet principle \cite{born_principles_1999}, and as such there is still a Kramers-like double degeneracy across the Brillouin zone. In the next section we will directly investigate how the reduction in crystalline symmetry manifests in the topological behavior. 

\subsection{Chern Number and Berry Curvature of the Rhombic EM Cell}

To determine the topological behavior of the unit cell, we first note that full behavior of the system includes the effects of both the electric and magnetic fields, and as such would ordinarily require vectorial descriptions of both. However, as shown in \cite{khanikaev_photonic_2013}, it is sufficient to consider the hybrid combination of the $E$ and $H$ components directed out-of-plane, which describe the pseudo-spin projected modes $\psi^\pm = \sqrt{\epsilon_0}E_z\pm\sqrt{\mu_0}H_z$. As we shall see, we may also investigate the $E_z$ and $H_z$ field components separately to observe how the reduction in crystalline symmetry manifest as well. 

For a $C_{n}$-symmetric system, it is possible to evaluate the spin Chern number via computation of the rotation eigenvalues of the spin-projected eigenfields at different high symmetry points (HSPs) within the Brillouin zone \cite{chen_manipulating_2015}. However, as we are interested in seeing the effects of these symmetry alterations, it is more illuminating to instead work with the Berry curvature \cite{fukui_chern_2005} and Wilson loop spectra \cite{alexandradinata_wilson-loop_2014}. Here we employ the numerical tools from \cite{bisharat_photonic_2021} for the topological calculations and full-wave simulated eigenfields computed within Ansys HFSS. 

\begin{figure}
	\centering
	\includegraphics[width=\linewidth]{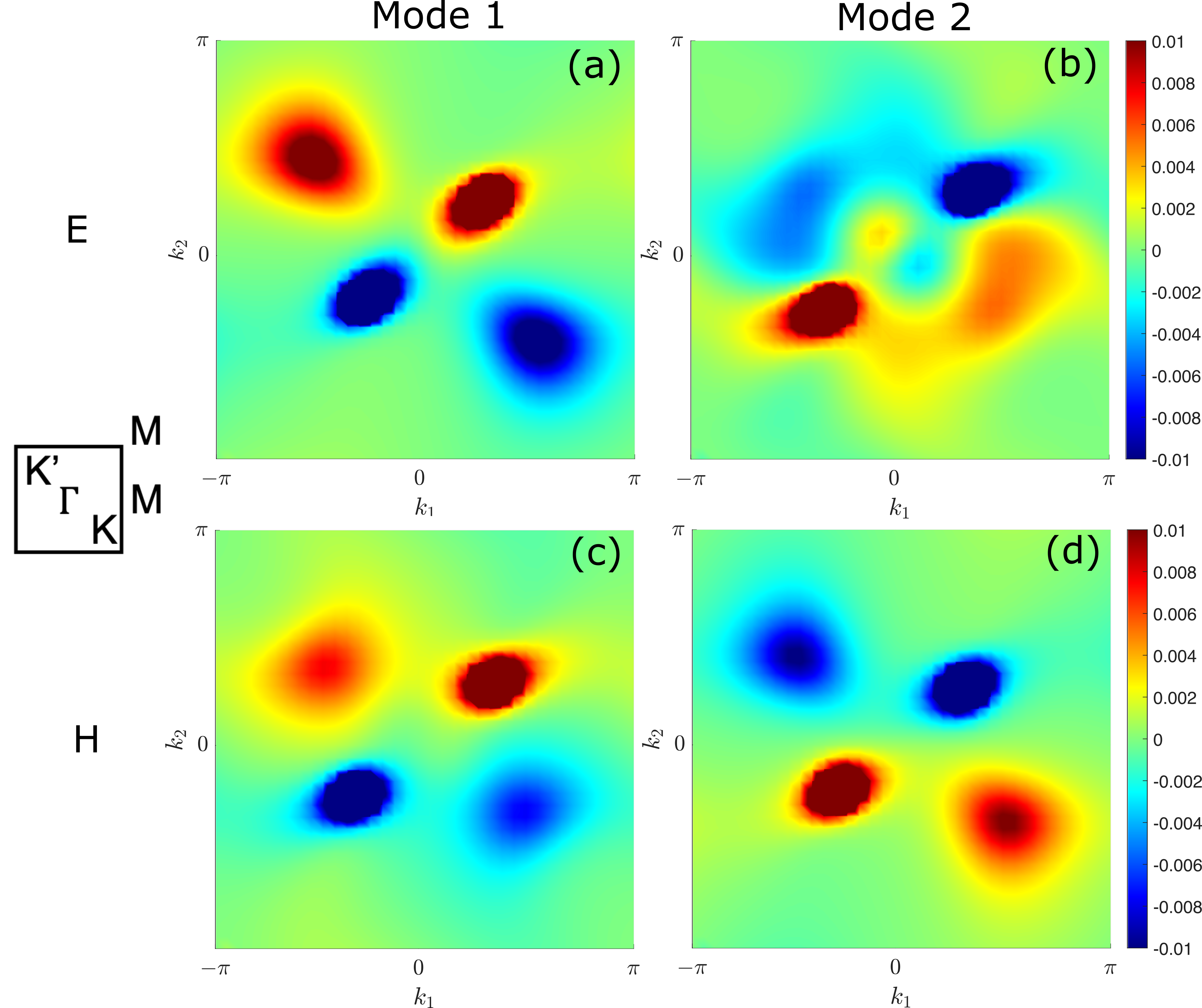}
	\caption{Berry curvatures of the rhombic cells. (a) and (b) show the Berry curvature of the $E_z$ fields considered in isolation for the first and second modes below the bandgap, respectively, while (c) and (d) show the Berry curvature for the $H_z$ fields in isolation for the same first and second modes, respectively. The Brillouin zone boundaries are illustrated in the left inset. The sharply defined peaks near the $\Gamma$ point correspond to degeneracies forced by the boundary conditions, while the diffuse peaks near the $K/K'$ points arise from the bianisotropic coupling of the $E$ and $H$ fields.}
	\label{fig:rhomCurv}
\end{figure}

We first compute the Berry curvature for the $E_z$ and $H_z$ components of the eigenmodes separately, the results of which are shown in Fig. \ref{fig:rhomCurv}. The lowest two eigenmodes are considered, and the calculation is performed over a uniform square grid $k_{i} = [-\pi,\pi]$ for $i=[1,2]$, along the two lattice vectors. In the simulation, the top and bottom surfaces are set to perfect magnetic insulation boundary conditions, which remove the components of the modes that exist above the light cone, purely for the sake of visualization and numerical stability.

In both the $E$ field and $H$ curvature distributions there are four distinct peaks, two localized near the $K/K'$ points, and two at interior points of the BZ. The former represent the topologically nontrivial mixing of the two field components, while the latter are a result of the boundary conditions imposed on TE and TM surface waves. Namely, the TM mode (in the absence of the numerically imposed magnetic boundary conditions) extends all the way to zero frequency at the $\Gamma$ point, while the TE mode has a finite cutoff frequency \cite{solymar_waves_2009}. This results in an "accidental" degeneracy which is not relevant to the topology of the model, occurring at the TE mode cutoff. Regardless of the exact cutoff, this results in spikes in the curvature of equal and opposite sign which integrate to zero. See also Appendix \ref{app:gauge} which discusses other numerical effects of the computation.

The more gradual accumulation of curvature near the $K/K'$ point, on the other hand, is directly related to the quantum spin Hall effect being emulated in the model. We note that the $E$ and $H$ field distributions of those points are not perfectly equal, and both have a finite spread over a wider area across the BZ. This is in contrast to the hexagonal case of Ref. \cite{bisharat_electromagnetic-dual_2019}, which has sharp features tightly localized near the $K/K'$ points, as they are a close analogue to the Dirac equation, as shown in Appendix. \ref{app:hex}. 

We will note that a non-zero accumulation of curvature localized around the $K/K'$ in valley-Hall phase systems leads to nontrivial edge states \cite{xiao_valley-contrasting_2007}, provided the two valleys are well separated in reciprocal space. Within the electromagnetic system, the integration of these curvatures directly does not lead to an ideally quantized valley Chern number $2*C_{valley} \notin \mathbb{Z}$ \cite{yang_evolution_2021}, and the topology is more accurately assessed by the sign different of the two valleys. In the $C_{2h}$ unit cell studied here the propagation direction is still along the $K/K'$ direction regardless, and so we might naively consider the system to be within a valley-like phase. However, the curvature distributions shown in Fig. \ref{fig:rhomCurv} have two key differences from a valley-structure: they are hybridized modes, and there are two separate, partially degenerate modes below the bandgap. Hence, we must instead consider the pseudo-spin fields to fully capture the effect. 

\begin{figure}
	\centering
	\includegraphics[width=\linewidth]{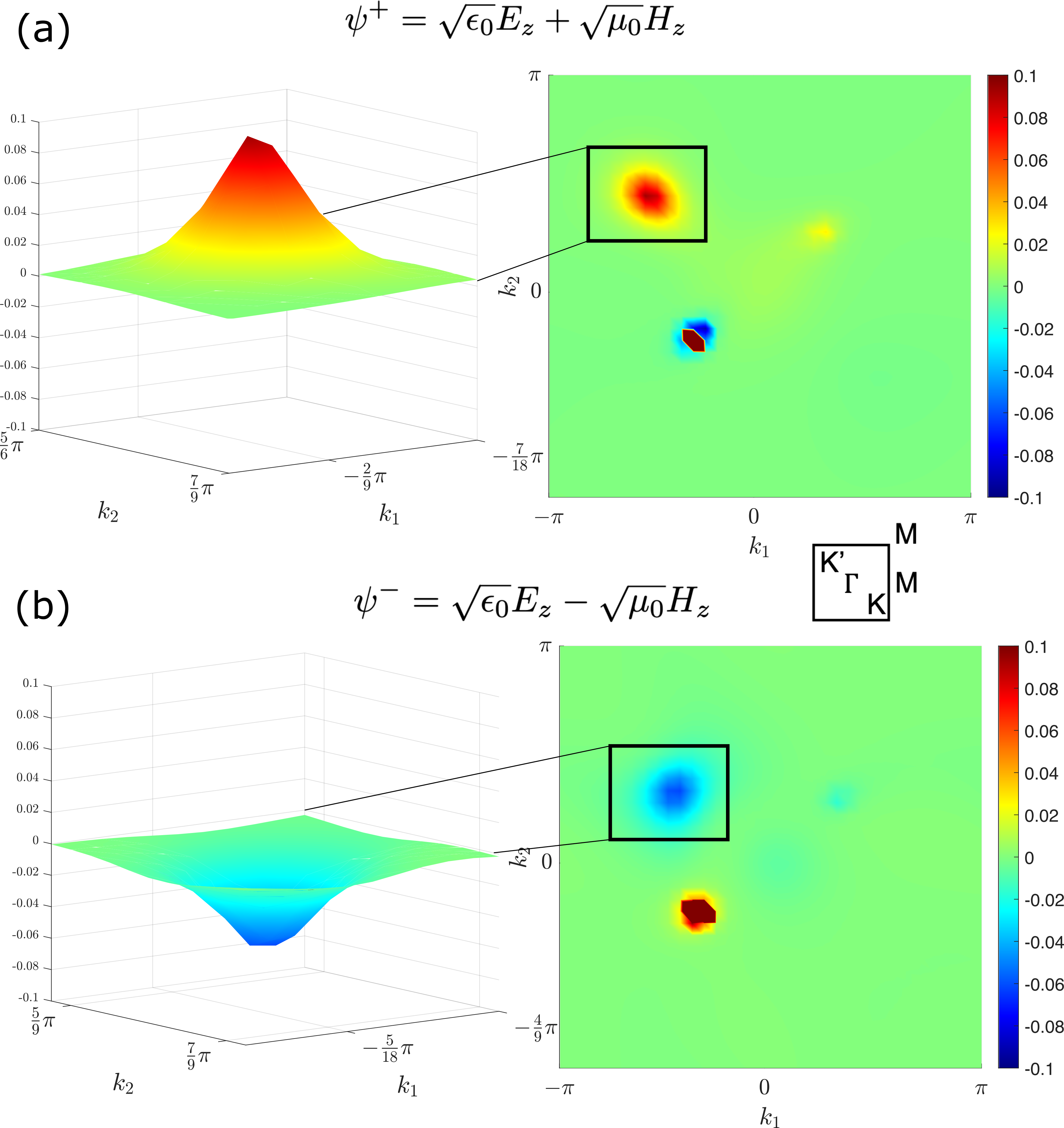}
	\caption{Spin-projected Berry curvatures of the rhombic cells. (a) shows the Berry curvature of the first mode projected into the (pseudo)spin-up space of $\psi^+ = \sqrt{\epsilon_0}E_z + \sqrt{\mu_0} H_z$, with the resulting positive accumulation concentrated near $K'$. (b) shows the Berry curvature of the second mode projected into the (pseudo)spin-down space of $\psi^- = \sqrt{\epsilon_0}E_z - \sqrt{\mu_0} H_z$, with the resulting negative accumulation concentrated near $K'$. Inset to the left are perspective zoom-in of the textures near the peak. }
	\label{fig:rhomSpin}
\end{figure}

To do so, we repeat the Berry curvature calculation in Fig. \ref{fig:rhomSpin}, only here we project the first and second modes into the (a) spin up ($\psi^+ = \sqrt{\epsilon_0}E_z + \sqrt{\mu_0}H_z$) and (b) spin down ($\psi^- = \sqrt{\epsilon_0}E_z - \sqrt{\mu_0}H_z$) spaces. Several features are visible:
\begin{enumerate}
	\item There is only one accumulation of curvature, here localized at $K'$, with zero at $K$. This is the result of projecting only one mode into each spin subspace. In the electronic equivalent plot, there is only one curvature, as the Kramers degenerate pairing of the two spin states requires the non-Abelian curvature \cite{wilczek_appearance_1984}. Hence, in this calculation we are selectively removing half of the modes, each of which contribute half of the curvature accumulation, one at $K$, the other at $K'$. An example of this non-Abelian calculation for the hexagonal 2-band manifold is shown in Appendix \ref{app:nonAbel}. 
	\item The sharpness of the peaks is greater than those of Fig. \ref{fig:rhomCurv}. This is a result of the bianisotropically-induced hybridization being selectively higher closer to the $K/K'$ points themselves. The absolute magnitude is likewise increased, seen on the color bar maxima. 
	\item The accidental degeneracies at the interior points of the Brillouin zone imperfectly cancel out. The simulation grid has a plaquette spacing of 37x37, and as such the sharply defined peaks are not exactly aligned numerically. However, they shrink in magnitude, and still do not contribute to the overall integrated value. 
\end{enumerate}

Critically, Fig. \ref{fig:rhomSpin} shows that the same pseudospin topology of the original hexagonal lattice (also given in Appendix \ref{app:hex}) is retained in the rhombic structure, despite the lowering of the point group symmetry. 

\subsection{Wilson Loop Spectra of the EM Rhombic Cell}

\begin{figure*}
	\centering
	\includegraphics[width=\linewidth]{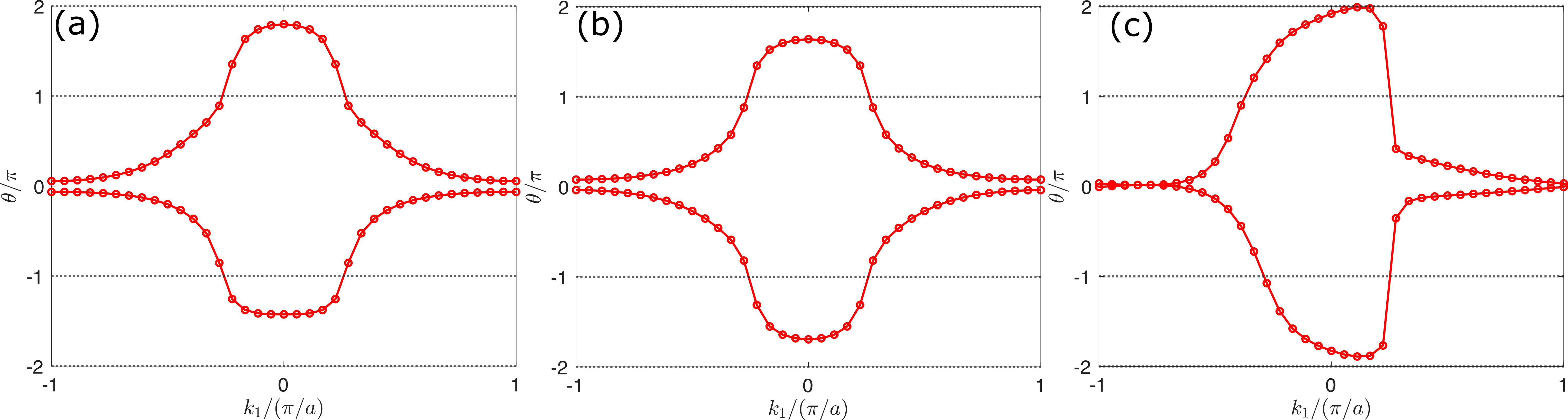}
	\caption{Wilson loop spectra of the rhombic cells. (a) and (b) show the Wilson loops for the non-Abelian calculation of the first and second bands using only the $E_z$ and $H_z$ eigenfields, respectively. (c) shows the non-Abelian spectra of the first and second bands using the spin-projected eigenfields $\psi^\pm = \sqrt{\epsilon_0}E_z \pm \sqrt{\mu_0}H_z$.}
	\label{fig:EMWilson}
\end{figure*}

The ease of separability of the two modes within the EM cell lends itself to the Berry curvature computation, as several key features can be visually identified without further calculations. This is not the case within electronic systems, where the doubly degenerate nature of the bands makes the winding of the Wilson loop eigenspectra a better tool for analyzing the topology \cite{alexandradinata_wilson-loop_2014, wang_band_2019}. The Wilson loop is also well-suited to handle the numerical artifacts present within the computation, as we shall see. The methods of \cite{bisharat_photonic_2021} can also be employed to compute this for the present model, as shown in Fig. \ref{fig:EMWilson}. 

In Fig. \ref{fig:EMWilson}(a) and (b), the Wilson loop is computed for the partially degenerate manifold of the lowest two modes for the $E_z$ and $H_z$ components, respectively. The $2\pi$ periodicity of the Brillouin zone is here extended, as indicated by the dashed lines, to visualize the smooth evolution of the two bands. These results parallel that of Fig. \ref{fig:rhomCurv}. We observe that both the $E$ and $H$ components indicate a non-trivial winding, as they, taken together, wind the full $2\pi$, representing an obstruction to creating localized Wannier functions \cite{paz_tutorial_2020}. The time-reversal nature of both components is also visible by the symmetry across the $k_1 = 0$ point for both eigenfields. 

However, as mentioned previously the two field components are not inseparable physically, the modes being fully hybridized. Hence, we again compute the Wilson loop via the spin-projected eigenfields $\psi^\pm = \sqrt{\epsilon_0}E_z\pm\sqrt{\mu_0}H_z$, shown in Fig. \ref{fig:EMWilson}(c). Here we observe that there is still a non-trivial winding, but the symmetry about the $k_1=0$ point is broken. This recalls the fact that we are projecting into the pseudo-spin space, both of which separately act like time reversal broken (Chern-like) systems, and so are not constrained to the same mirror symmetry about $k_1=0$.

\section{Hamiltonian Model and Symmetry Breaking}\label{sec:ham}
The previous sections have examined the behavior of the specific EM dual unit cell numerically, and we have seen the non trivial topology appear. However, to see the generality of these results it is informative to consider a pared down model that can capture most of the symmetry. The simplest choice is the original Kane-Mele Hamiltonian, the tight binding description given by

\begin{equation}
	H_{KM} = \sum_{\langle i,j\rangle \alpha} t_1 c^\dagger_{i\alpha}c_{j\alpha} + \sum_{\langle\langle ij\rangle\rangle \alpha\beta}it_2 \nu_{ij}s^z_{\alpha\beta} c^\dagger_{i\alpha}c_{j\beta},\label{eq:KMH}
\end{equation}
with the first summation covering hopping from nearest neighbor (1NN) atoms, and the second sum covering second nearest neighbor (2NN) hopping between spins. $t_1$ denotes the hopping amplitude between atomic sites $i$ and $j$ with spin $\alpha$, $t_2$ is the 2NN hopping amplitude, $\nu_{ij}$ is the spin-dependent term that selects between spin states depending on direction, and $s^z_{\alpha\beta}$ is the Pauli matrix for the spin space. 

\begin{figure}
	\centering
	\includegraphics[width=\linewidth]{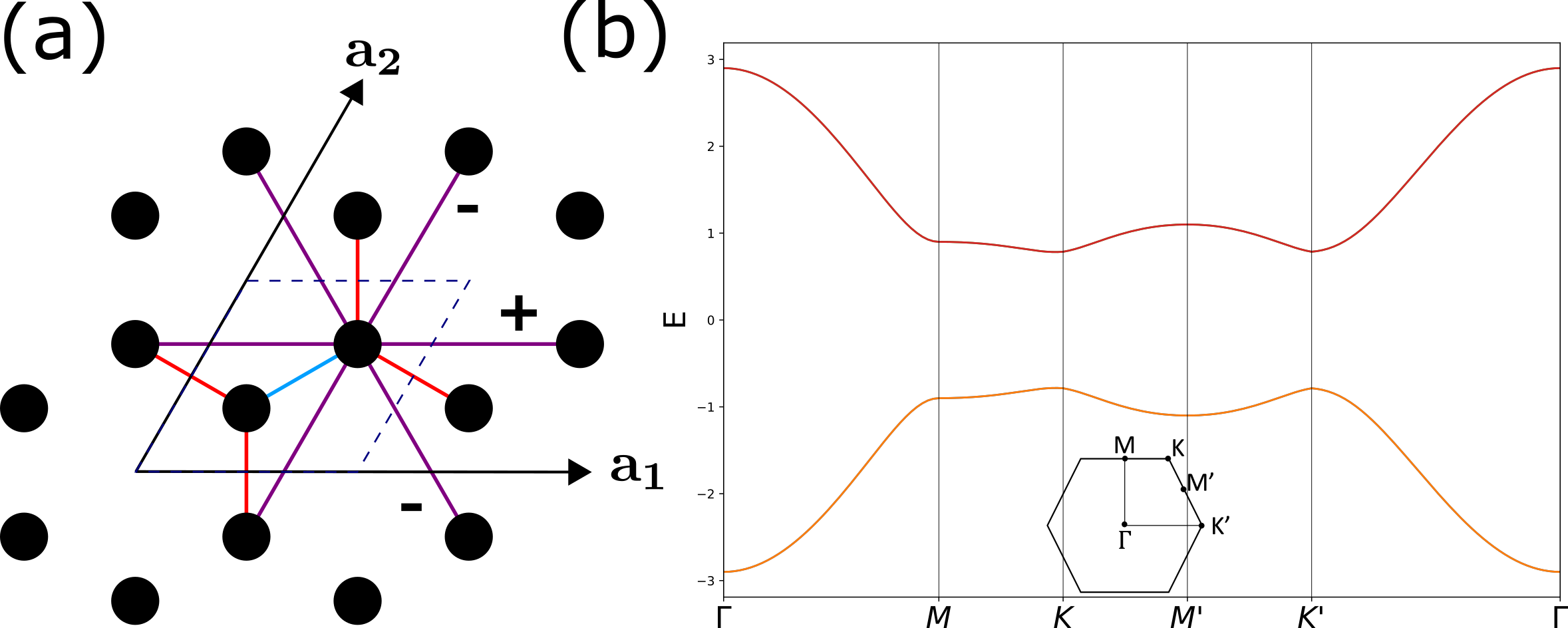}
	\caption[Band structure of the ]{Modified $C_{2v}$ KMH Model. (a) Unit cell and hopping arrangement of the modified Kane-Mele Hamiltonian. The unit cell is shown in the dashed blue rhombus, with two atomic sites inside. The red and blue lines denote the first nearest neighbor hopping terms. The red bonds are equal, but when the red bonds do not equal the blue bond, the model drops to $C_{2v}$ symmetry. The spin-dependent imaginary second nearest neighbor hopping terms are shown in the purple bonds, with only one site shown for clarity. (b) Representative band structure for the modified Kane-Mele Hamiltonian, Eq. \eqref{eq:C2KMH}. The 2-band TB model used has $t_1=1$ for intercellular hopping terms (red bonds in (a)), $t_1=0.9$ for intracellular hopping, $t_2=0.15$, and the spin texture as shown in (a). Note that the spin bands are doubly degenerate at all momenta, and the broken $C_{3}$ symmetry causes the $M$ points to differ. }
	\label{fig:KMHbands}
\end{figure}

To connect the model of Fig. \ref{fig:rhomCell}(a) to $H_{KM}$, we will need to appropriately alter the crystallographic properties to match. In Eq. \eqref{eq:KMH}, the first sum has equal hopping amplitudes among first nearest neighbor lattice sites (e.g., between the two sublattices). Before the 2NN terms are added, we have a simple model of graphene, which has point group $C_{6v}$, naturally possessing an enforced degeneracy at the $K$ point due to the existence of 2D irreps \cite{malterre_symmetry_2011}. To force this into $C_{2v}$, we can simply alter one of the three 1NN bonds, which preserves the two mirror symmetries. This hopping texture is easier to see via the real space arrangement, which is shown in Fig. \ref{fig:KMHbands}(a). The single altered 1NN bond is shown in the light blue, and the 2NN spin-hopping terms (purple) are included as well. Note that even without the 2NN terms the model has a continuous bandgap, as expected. 

This modified model, given as
\begin{equation}
	H_{KM}^{C_{2v}} = \sum_{\langle i,j\rangle \alpha} t^{r(b)}_{1} c^\dagger_{i\alpha}c_{j\alpha} + \sum_{\langle\langle ij\rangle\rangle \alpha\beta}it_2 \nu_{ij}s^z_{\alpha\beta} c^\dagger_{i\alpha}c_{j\beta},\label{eq:C2KMH}
\end{equation}
includes the site-dependent 1NN hopping amplitude $t_1^{a(b)}$, where we define $t_1^a$ to be the hopping between the red bonds in Fig. \ref{fig:KMHbands}(a) (e.g., intercellular terms) and $t_1^b$ to be the hopping between the blue bonds in Fig. \ref{fig:KMHbands}(a) (e.g., intracellular terms). In the case where $t_1^a=t_1^b$, Eq. \ref{eq:C2KMH} reduces back to the original KMH of Eq. \ref{eq:KMH}. 

In a lattice model, Eq. \eqref{eq:C2KMH} can be represented by the Bloch Hamiltonian $H(\mathbf{k},t_1^r,t_1^b,t_2)$ as 
\begin{equation}\label{eq:bloch}
	H(\mathbf{k},t_1^r,t_1^b,t_2) =
	\begin{pmatrix}
		H^+(\mathbf{k},t_1^r,t_1^b) & t_2H_\Delta^\dagger \\
		 t_2H_\Delta & H^-(\mathbf{k},t_1^r,t_1^b)
	\end{pmatrix}
\end{equation}
where the Hamiltonian for a single spin is given by the graphene-like two band model $H^\pm$ given in Eq. \ref{eq:Hpm}.

\begin{widetext}
\begin{equation}\label{eq:Hpm}
	H^\pm(\mathbf{k},t_1^r,t_1^b,\delta) = 
	\begin{pmatrix}
		\delta & t_1^{b*} + t_1^{r*}(e^{ik_x} + e^{i(\frac{1}{2}k_x +\frac{\sqrt{3}}{2}k_y)}) \\
		t_1^b + t_1^r (e^{-ik_x} + e^{-i(\frac{1}{2}k_x +\frac{\sqrt{3}}{2}k_y)}) & \delta
	\end{pmatrix},
\end{equation}
\end{widetext}

In Eq. \eqref{eq:bloch}, we have chosen real space lattice vectors $\mathbf{a_1} = (1,0), a_2 = (\frac{1}{2},\frac{\sqrt{3}}{2})$. On-site potential terms are given as $\delta$. The spin-coupling terms $H_\Delta$ are given as 

\begin{widetext}
\begin{equation}
	H_\Delta(\mathbf{k}) = 
	\begin{pmatrix}
		-ie^{i k_x} + ie^{i(\frac{1}{2}k_x +\frac{\sqrt{3}}{2}k_y)} - ie^{i(\frac{1}{2}k_x -\frac{\sqrt{3}}{2}k_y)}& 0\\
		0 & ie^{i k_x} - ie^{i(\frac{1}{2}k_x +\frac{\sqrt{3}}{2}k_y)} + ie^{i(\frac{1}{2}k_x -\frac{\sqrt{3}}{2}k_y)}
	\end{pmatrix},
\end{equation}
\end{widetext}

A representative tight binding band structure is shown in Fig. \ref{fig:KMHbands}(b). Note that in the computation we have set the on-site potential terms to zero ($\delta = 0$), and as such the two spin bands above and below the bandgap are fully degenerate at all $\mathbf{k}$. With $t_1^r \ne t_1^b$, the $C_{2v}$ symmetry results in an expanded Brillouin zone path, as the $M$ and $M'$ points are no longer fixed to be equal. Note that in Fig. \ref{fig:rhomCell}(b) the path excludes the $M'$ point, as it does not contribute to later analysis, but is shown here for completeness in the TB band structure. The general shape is in general good agreement with Fig. \ref{fig:rhomCell}(b), with the only notable difference being the slight band repulsion observed in Fig. \ref{fig:rhomCell}(b). This arises due to minor deviations between the local field structure interacting with the metal surfaces, which can be mimicked in the TB model by a Rashba-like term (see Appendix \ref{app:rashba}), but this does not alter the underlying topology \cite{kane_quantum_2005}. 

\begin{figure}
	\centering
	\includegraphics[width=\linewidth]{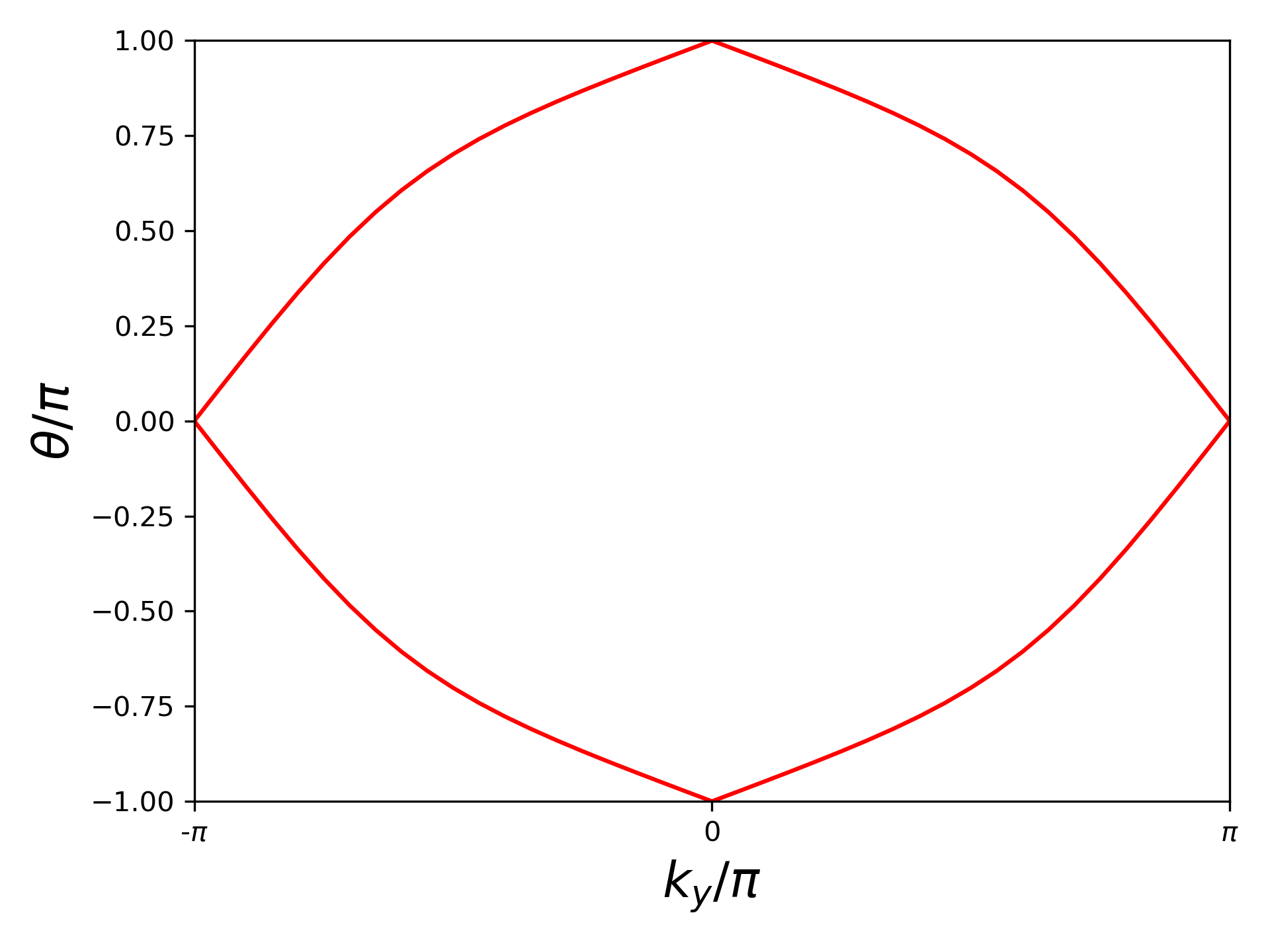}
	\caption[Wilson loop spectra of the $C_{2v}$ KMH]{The Wilson loop spectra of the modified $C_{2v}$-based Kane-Mele Model of Eq. \eqref{eq:C2KMH} for the lower two spin-degenerate states. Here the non-Abelian calculation of the Wilson loop is performed. }
	\label{fig:KMHWilson}
\end{figure}

In the fully Kramers degenerate version, the Berry phase is not a useful quantity. To check the topological properties of this reduced symmetry model, we can instead compute the Wilson loop spectra, which is shown in Fig. \ref{fig:KMHWilson}(a). We see that the two spin bands wind oppositely, and cover the full BZ range, indicating a $\mathbb{Z}_2 = 1$, identical to the standard Kane-Mele result. For this result, the 1NN hopping is set to $t_1^r = 1, t_1^b = 0.9$, and spin-orbit coupling strength $t_2 = 0.14$. This is in close agreement with the numerical Wilson loop from the EM unit cell shown in Fig. \ref{fig:EMWilson}, and naturally agrees with the Kane-Mele result. 

\begin{figure}
	\centering
	\includegraphics[width=\linewidth]{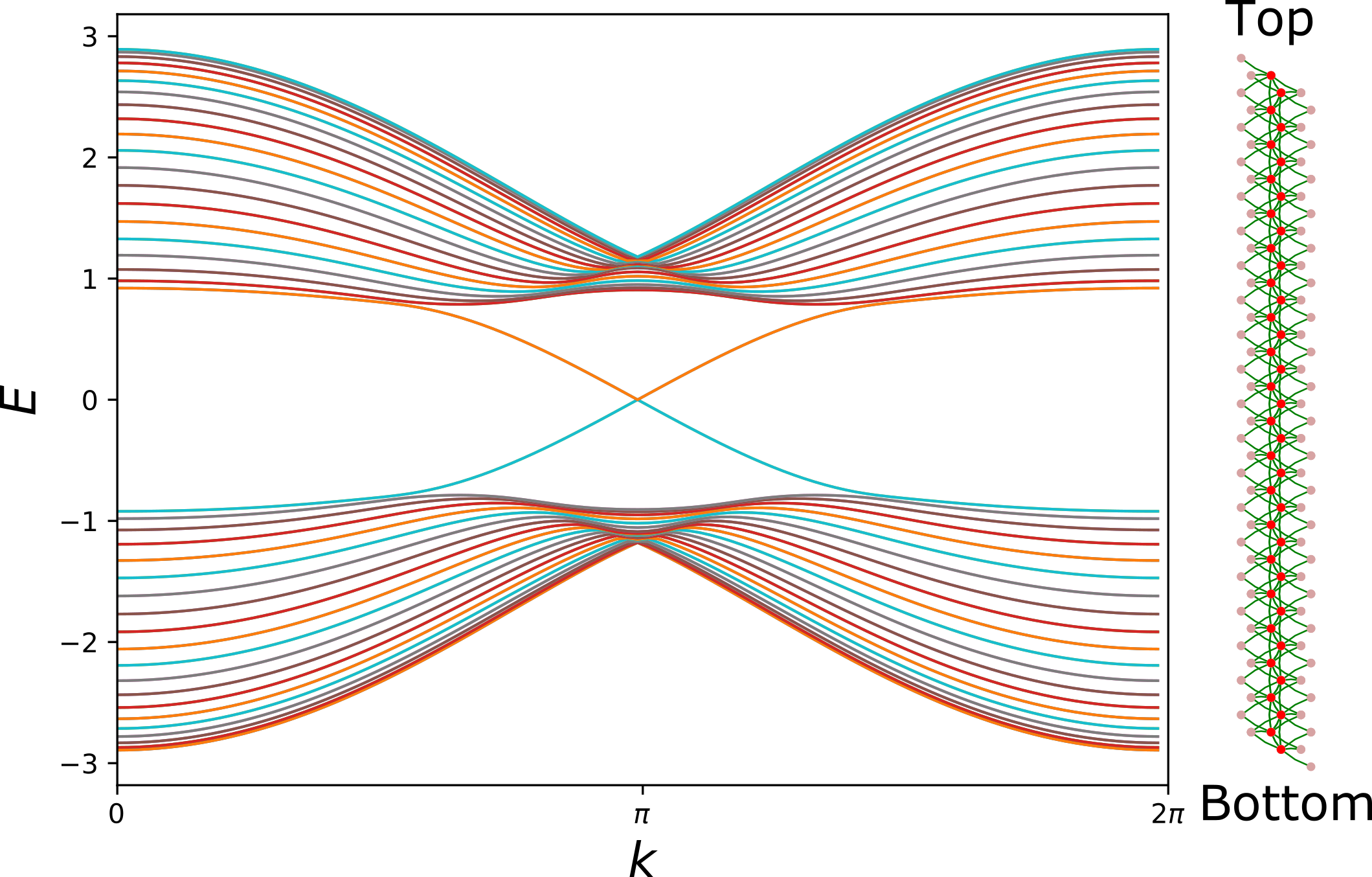}
	\caption[Surface bands for the $C_{2v}$ KMH]{Surface band structure for a finite ribbon of unit cells of Eq. \ref{eq:C2KMH}, using $t_1^r$ = 1.0,  $t_1^b$ = 0.9, and $t_2$ = 0.15}
	\label{fig:C2KMHribbon}
\end{figure}

To see the surface states appear on a finite edge, we can also compute the surface band structure for this model, which is given in Fig. \ref{fig:C2KMHribbon}. We see the expected doubly-degenerate states that propagate on the top and bottom finite surfaces of the ribbon, trapped within the bulk bandgap. 

From these, we can see that the reduction in symmetry does not destroy the expected topological insulator behavior. It is worth noting that the mapping from TB model and the electromagnetic model in Fig. \ref{fig:rhomCell} is not exact; the detailed energetics of the system differ, and the tight binding approximation used does not fully capture the small details that the PTI model possesses. Nevertheless, the fact that we can break the requisite symmetries in the idealized Kane-Mele model without destroying the expected topological surface states is indicative of the effect that switching from a hexagon to a rhombus has on the topology.

\section{Altering the Rhombic Angle}\label{sec:angle}
From the prior analysis we can see that the primary symmetry required for the nontrivial surface states to exist is EM duality, $\mu=\epsilon$. However, the question then arises on what role if any the crystalline symmetry has on the existence of the states. We can easily observe that adiabatically altering the lattice angle within the simplified TB model of Eq. \ref{eq:C2KMH} does not close the bandgap, and therefore does not change the topology. However, as noted previously the analogy between the TB model and the EM unit cell is not exact, and we must instead evaluate the topological properties of the duality cell itself. 

\begin{figure}
	\centering
	\includegraphics[width=\linewidth]{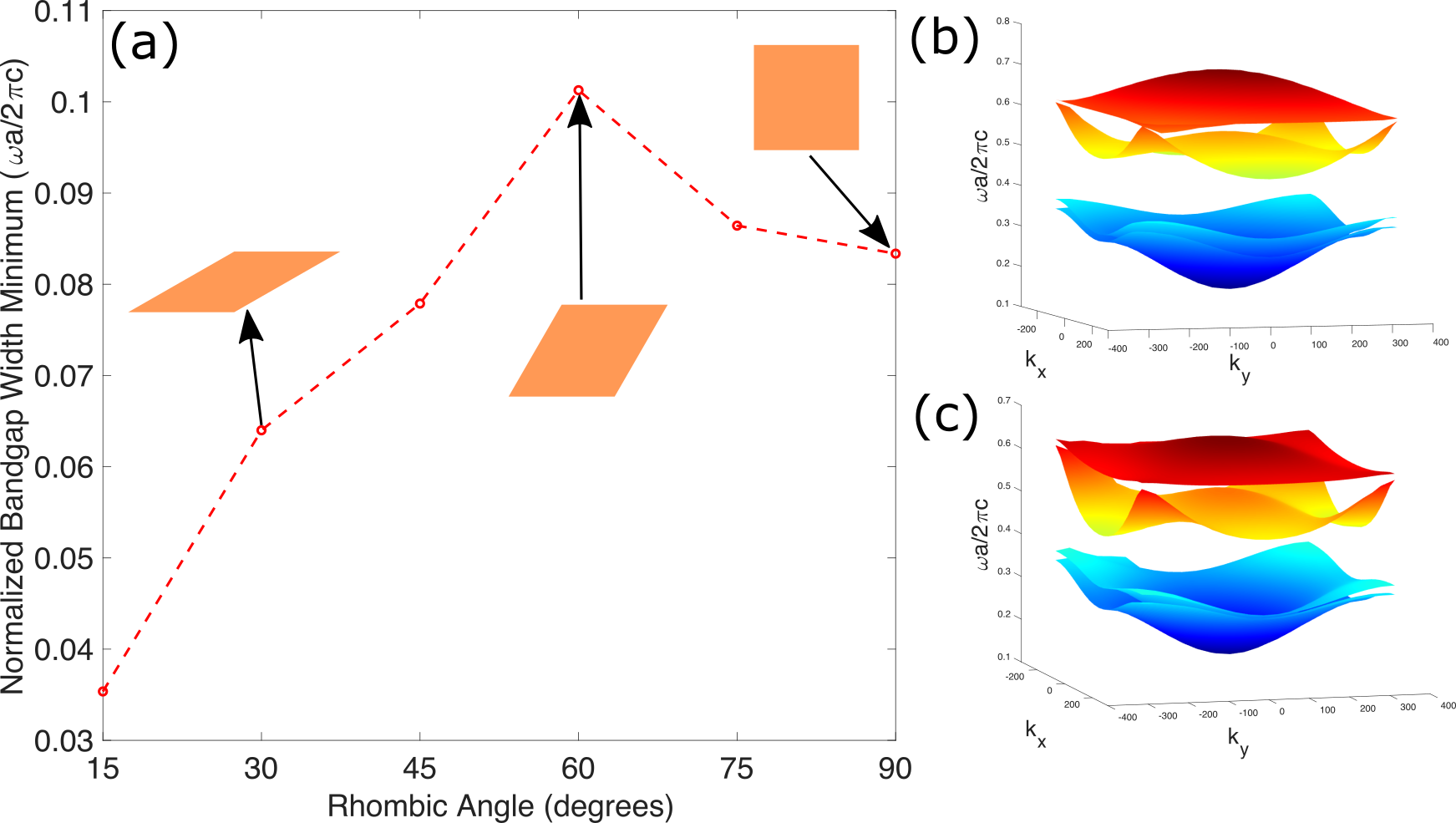}
	\caption{(a) Normalized bandgap width as a function of rhombic angle. Here 60 degrees corresponds to the rhombic model shown in Fig. \ref{fig:rhomCell}. Inset are the 30, 60, and 90 degree cell geometries. (b) and (c) show the 3D bandstructures for the 60 degree (rhombic) and 90 degree (square) lattices, respectively.}
	\label{fig:angles}
\end{figure}

To see how far we can push the basic rhombic unit cell, we can alter the angle $\theta$ from the nominal 60 degrees (which lives on the triangular lattice) to any arbitrary angle. This includes the square lattice of $\theta=90$ degrees. The alteration of symmetry implies differing properties for the Berry curvature accumulation, but we may predict that a small change $\delta \theta$ does not immediately destroy the edge states, as the bandgap remains open. To see this directly, Fig. \ref{fig:angles}(a) shows the minimum bandgap size of the EM cell from Fig. \ref{fig:rhomCell}(a) as the angle $\theta$ is swept. We notice that as the angle approaches zero and the cell becomes more oblong, the bandgap starts to close, and likewise as it pushes beyond 60 degrees. The 60 degree point maximizes the bandgap width, which may be beneficial for applications that require large bandwidth. Note that to properly measure the gap width, the eigenstates are simulated across the full BZ, two examples of which are shown in Fig. \ref{fig:angles}(b-c) for the 60 and 90 degree unit cells. 

While the bandgap remains open, propagating edge states are not guaranteed to be found, as the pseudospin definitions of the topological modes start to mix as the angle becomes more extreme. At angles far enough from the triangular lattice, the $\psi^\pm$ behavior is no longer a good description of the hybrid behavior, and the bulk topological signature gets suppressed, unlike in the electronic version. This can be seen directly in the 90 degree (square) case, where we effectively have a duality spin version of the "line wave" concept of Ref. \cite{bisharat_guiding_2017}, but here we are considering the bandgap region, rather than the bulk-mode region of the line wave concept. The topological analysis is repeated for this configuration in Appendix \ref{app:square}, where we see the non-trivial signatures disappear.

\section{Device Implementations}\label{sec:dev}
Beyond the topological features observed in reciprocal space, there is a practical benefit to the rhombic unit cell as presented: it is straightforward to construct robust waveguides from them. In Ref \cite{davis_classical--topological_2021}  a coupler design was presented which required carefully adjusting the field match at the connection point between the classical and topological waveguiding regions. This was complicated by the hexagonal unit cell, which moves laterally as you move along the propagation ( "zig-zag") path. In the rhombic case, the propagation path is a straight line, significantly easing this difficulty. This may also prove useful for topological-based antenna designs \cite{singh_topological_2021}, which has recently been shown directly \cite{ahmad_abtahi_realizable_2024}.

\begin{figure}
	\centering
	\includegraphics[width=\linewidth]{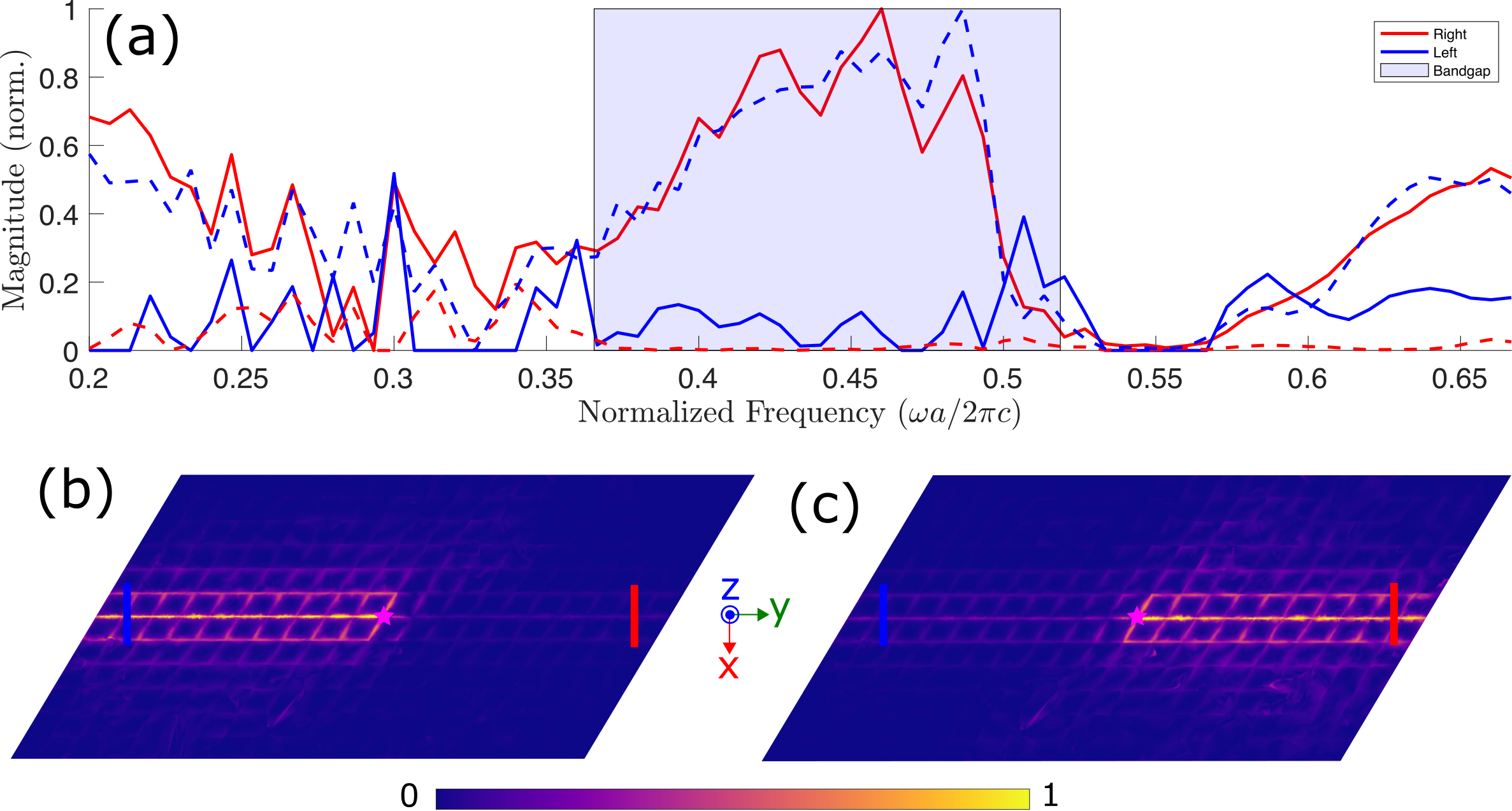}
	\caption{Spin-filtering property of the rhombic structure. (a) Normalized transmission spectra in the right (red) and left (blue) directions along a straight interfacial paths using the EM unit cell of Fig. \ref{fig:rhomCell}(a). Solid (dashed) lines show the case of the spin-up (spin-down) excitation using a Hertzian dipolar source of $E_z \pm Z_0H_z$, where positive is defined as spin-up and negative as spin-down. The bandgap of the unit cell is indicated in the shaded region. (b) and (c) show the magnitude of the electric fields in the middle of the structure at a normalized frequency of 0.45 for the spin down and up sources, respectively. The dipolar source is noted by a pink star, while the locations where the transmission is computed for (a) are indicated by the red and blue rectangles in (b) and (c).}
	\label{fig:rhomSource}
\end{figure}

An example of unidirectional excitation of the pseudospin states is shown in full-wave simulated result shown in Fig. \ref{fig:rhomSource}. The pink star on panels (b) and (c) denotes a hybrid source of electric and magnetic dipoles, the sign of which determines the propagation direction \cite{khanikaev_photonic_2013}. We see the mode is tightly confined, and negligible field propagates in the opposite direction. To see this more quantitatively, Fig. \ref{fig:rhomSource}(a) shows the power flow moving to the right (red lines) and left (blue lines). The dashed lines are for the case of the spin-down source in Fig. \ref{fig:rhomSource}(b), while the solid lines are for the spin-up source of Fig. \ref{fig:rhomSource}(c). The power flow is computed via the integral of the real part of the Poynting vector through a surface on either end of the simulation region, denoted by the red and blue rectangles in Fig. \ref{fig:rhomSource}(b) and (c). The boundaries of the simulated region are set to radiation boundary conditions, which largely remove reflections off the finite edges, although some back-reflection is observed in the result. In Fig. \ref{fig:rhomSource}(a) only the positive component of the integral is considered, to remove these reflections. The magnitudes for both directions are normalized to the peak for each direction of the source. 

We can see that in both cases there is substantially higher power flowing in one direction for frequencies within the bulk bandgap (denoted in the shaded region, which corresponds to the shaded region of Fig. \ref{fig:rhomCell}(b)) than the other. When the chirality of the source is flipped, we see the expected flipping of directions. In both configurations, there is much higher transmission through the chosen region within the topological bandgap vs the bulk region, as the bulk region spreads energy over much larger regions than the tightly-confined edge state. 

\begin{figure}
	\centering
	\includegraphics[width=\linewidth]{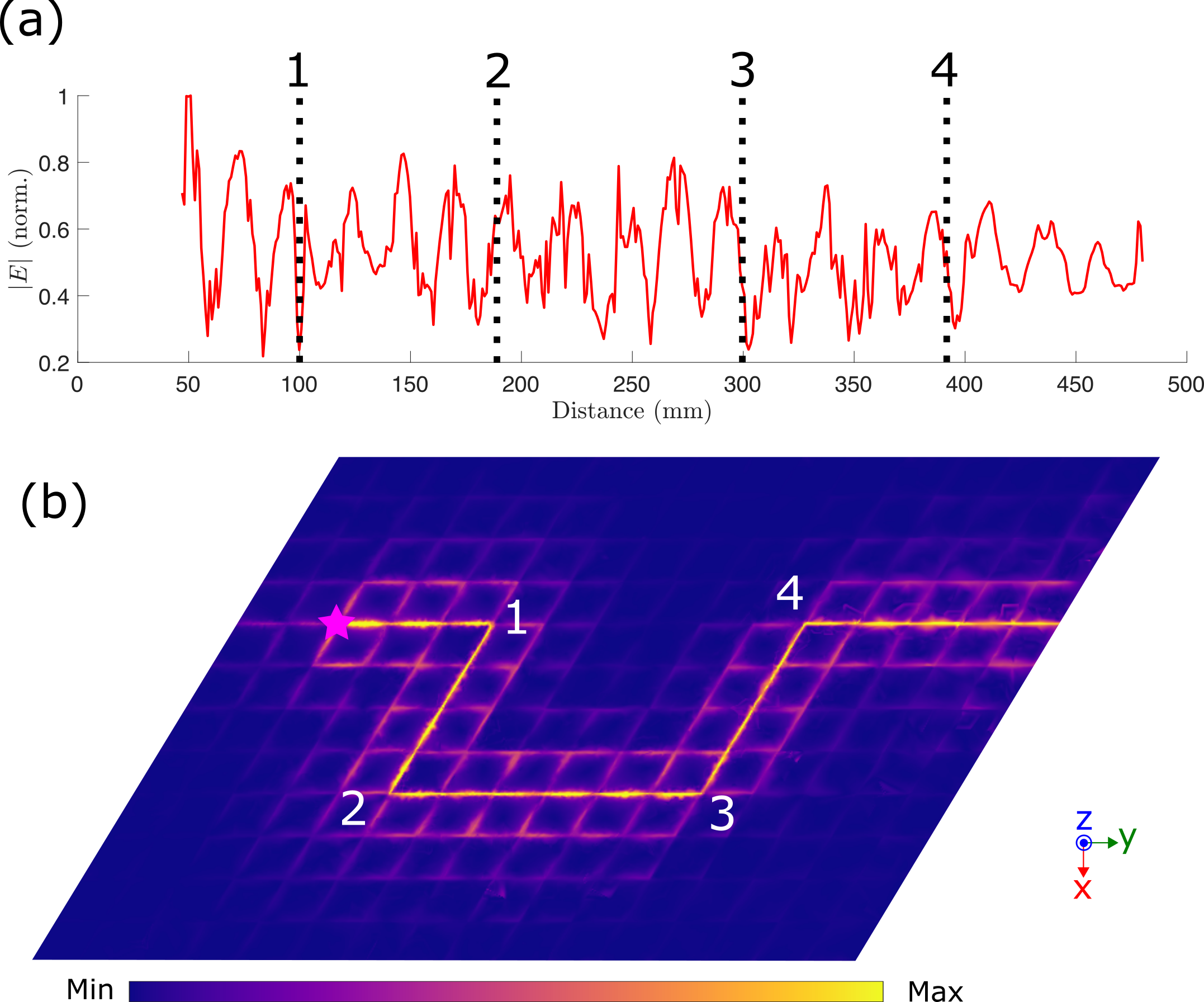}
	\caption{Demonstration of the spin modes' robustness to backscatter at sharp turns. (a) shows the magnitude of the electric field along the propagation path shown in (b). Four sharp turns are placed along the path, numbered 1-4, which do not indicate any reflections. The source is a spin-down $E_z - Z_0H_z$ at normalized frequency of 0.45, located at the pink star in (b). (b) shows the magnitude of the $E$ field in the $x-y$ plane taken at $z=0$.}
	\label{fig:rhomZag}
\end{figure}

Another hallmark feature of topologically robust waveguides are their robustness to reflections at sharp discontinuities \cite{ozawa_topological_2019}. To test this, another full-wave simulation is shown in Fig.  \ref{fig:rhomZag}, which includes several sharp bends, with the same spin-up polarized source used in Fig. \ref{fig:rhomSource}. In Fig. \ref{fig:rhomZag}(a) is shown the magnitude of the electric field along the propagation path, with the three sharp corners denoted by the numerals depicted in Fig. \ref{fig:rhomZag}(b). A small degree of propagation loss is observed (likely from evanescent leakage), but there is no noticeable reflection at any of the sharp turns themselves. 

\section{Conclusions}
In this paper we have demonstrated and analyzed a platform for studying photonic spin states in a rhombic lattice, which displays topological features. The nontrivial nature of the bulk material was computed via Berry curvature calculations from full-wave simulations, as well as via modifying the Kane-Mele Hamiltonian. The impact of altering the rhombic angle and relevant bandgap closing behavior was also presented. Finally, device implementations were shown, which provide a simple experimental platform useful for applications. 

\begin{acknowledgments}
This work was supported by AFOSR grant FA9550-21-1-0167. The authors would like to acknowledge fruitful discussion with S. Ahmad Abtahi during the preparation of this manuscript. 
\end{acknowledgments}

\appendix

\section{Hexagonal Unit Cells}\label{app:hex}

The Berry curvature for the $C_{6v}$ symmetric model studied in \cite{bisharat_electromagnetic-dual_2019} possesses two spikes at the $K/K'$ HSPs, each of which contribute a Berry phase of $\pm\pi/2$, or a spin Chern number of $\pm1/2$. The location of these points is fixed by the symmetry of point group: in the absence of bianisotropic coupling there is a Dirac cone pinned to the $K/K'$ points by the $C_{6v}$ rotational symmetry of the unit cells \cite{malterre_symmetry_2011}. The introduction of bianisotropy induces a bandgap, which causes the Berry curvature to accumulate in the region of the gapped Dirac crossing. Integration of the Berry curvature across the whole BZ yields a spin Chern number of $\pm 1$ for the upper and lower bands \cite{bisharat_photonic_2021}. 

\begin{figure}
	\centering
	\includegraphics[width=\linewidth]{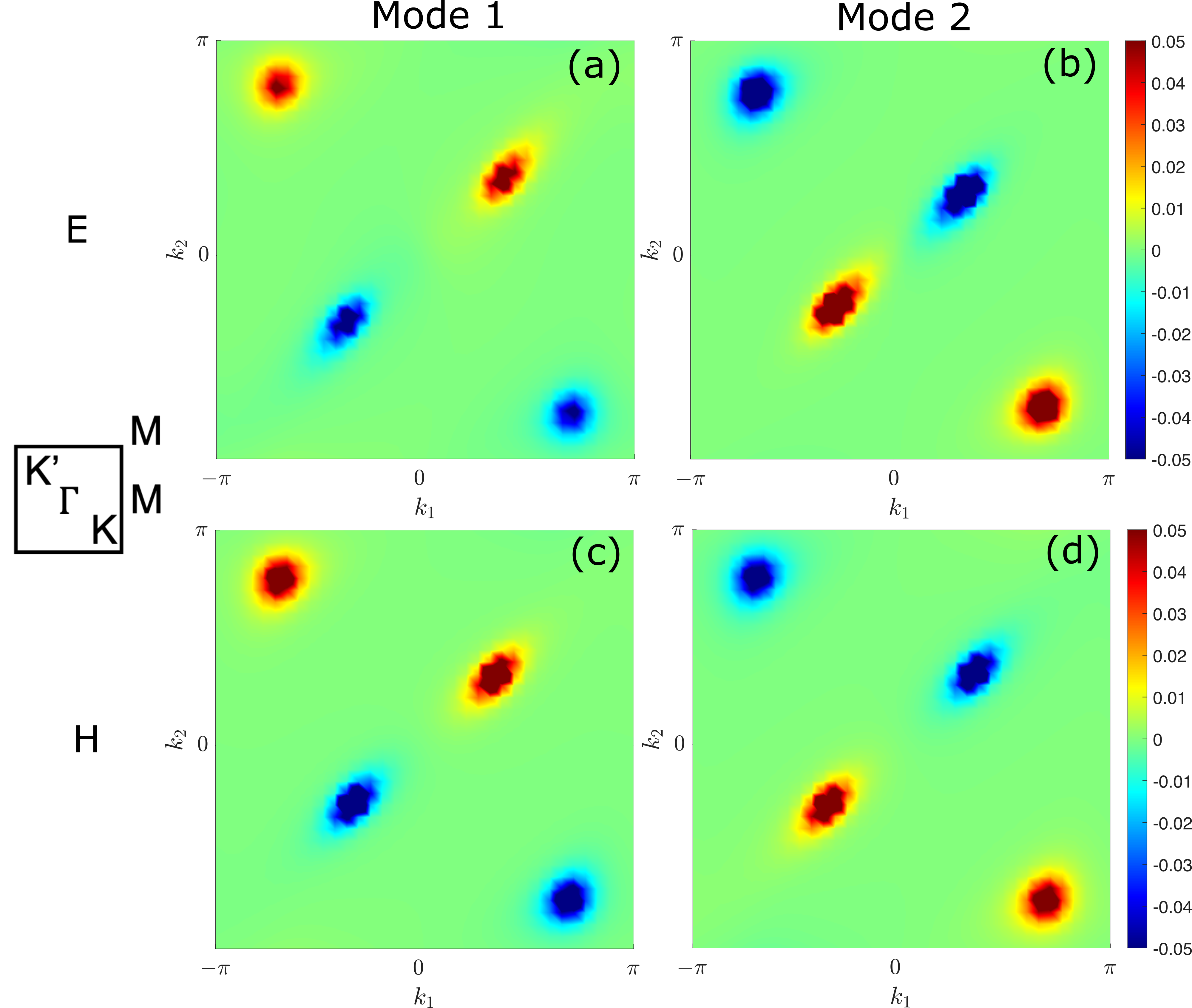}
	\caption{Berry curvatures of the hexagonal cells. The first row corresponds to the $E_z$ fields considered in isolation, while the second row corresponds to the $H_z$ fields in isolation. The first column is the first mode, and the second column is the second. Note that there is a symmetrical flip between the two.}
	\label{fig:hexCurv}
\end{figure}

To see this numerically, Fig. \ref{fig:hexCurv} shows the same 2-band simulation of the Berry curvature of the hexagonal unit cell as described in \cite{bisharat_electromagnetic-dual_2019}. Here, the same cell periodicity $a=20$ mm, thickness $h = 1.57$ mm, and frame width $b=0.43$ mm as that of the rhombic unit cell is used, with the only change being the choice of a hexagonal shape. As this considers only the $E_z$ and $H_z$ components separately, we again notice a valley-like distribution, with the same internal BZ spikes noted for the rhombic unit cell. The primary difference seen is that the peak is much more localized to the $K/K'$ point, forced there by the $C_{6v}$ symmetry. 

\begin{figure}
	\centering
	\includegraphics[width=\linewidth]{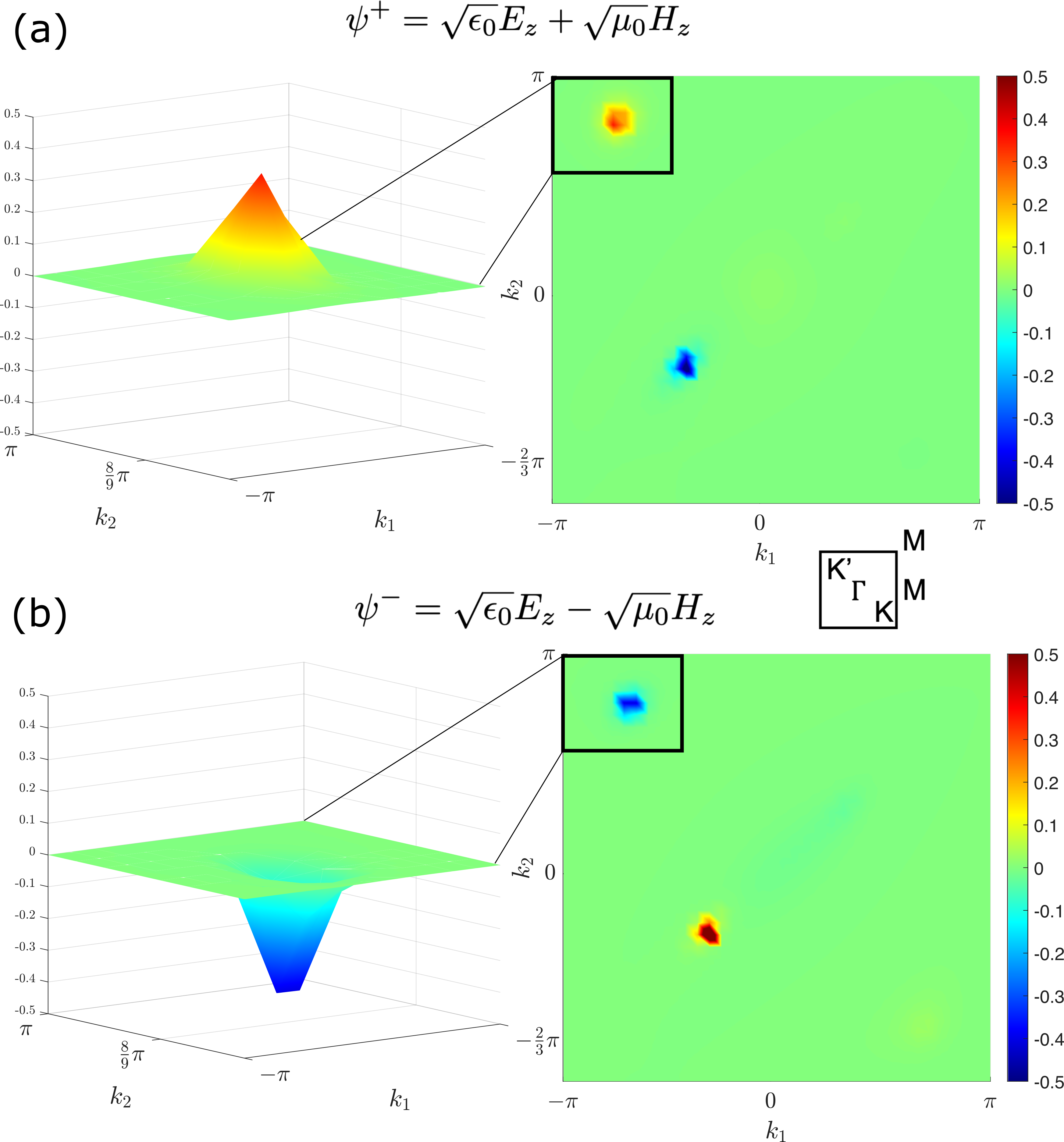}
	\caption{Berry curvatures of the hexagonal cells for the pseudospin projected eigenfields of $\psi^+ = \sqrt{\epsilon_0}E_z + \sqrt{\mu_0} H_z$, with the resulting positive accumulation concentrated near $K'$. (b) shows the Berry curvature of the second mode projected into the (pseudo)spin-down space of $\psi^- = \sqrt{\epsilon_0}E_z - \sqrt{\mu_0} H_z$, with the resulting negative accumulation concentrated near $K'$. Inset to the left are perspective zoom-in of the textures near the peak.}
	\label{fig:hexSpinCurv}
\end{figure}

Fig. \ref{fig:hexSpinCurv} shows the result of the Berry curvature when projecting into the spin-space. Once again we see the same localization effect, with clearly defined peaks pinned to $K'$, the sign of which is determined by the sign of the pseudospin definition. As shown in Appendix \ref{app:nonAbel}, the non-Abelian version of this calculation shows the full $K/K'$ mapping expected when considering both bands together projected on the same spin space. Integrating over each valley gives the expected $\pm\frac{1}{2}$. 

\begin{figure*}
	\centering
	\includegraphics[width=\linewidth]{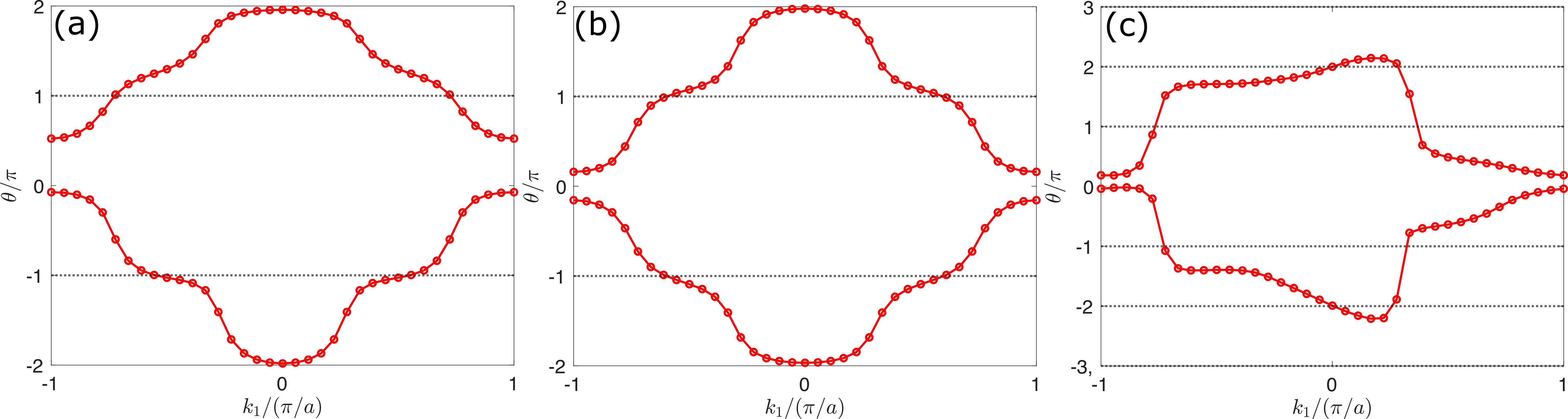}
	\caption{Wilson loop spectra of the hexagonal cells. (a) and (b) show the loops for the non-Abelian calculation of the first and second bands using only the $E_z$ and $H_z$ eigenfields, respectively. (c) shows the non-Abelian spectra of the first and second bands using the spin-projected eigenfields $\psi^\pm = \sqrt{\epsilon_0}E_z \pm \sqrt{\mu_0}H_z$.}
	\label{fig:hexEMWilson}
\end{figure*}

Finally, the Wilson loop spectra is given in Fig. \ref{fig:hexEMWilson}. We see the expected $2\pi$ winding, symmetry, and general features discussed in the main text, with the primary difference being the precise shape. 

\section{Non-Abelian Berry Curvature of the Spin-Projected Eigenfields}\label{app:nonAbel}

\begin{figure}
	\centering
	\includegraphics[width=\linewidth]{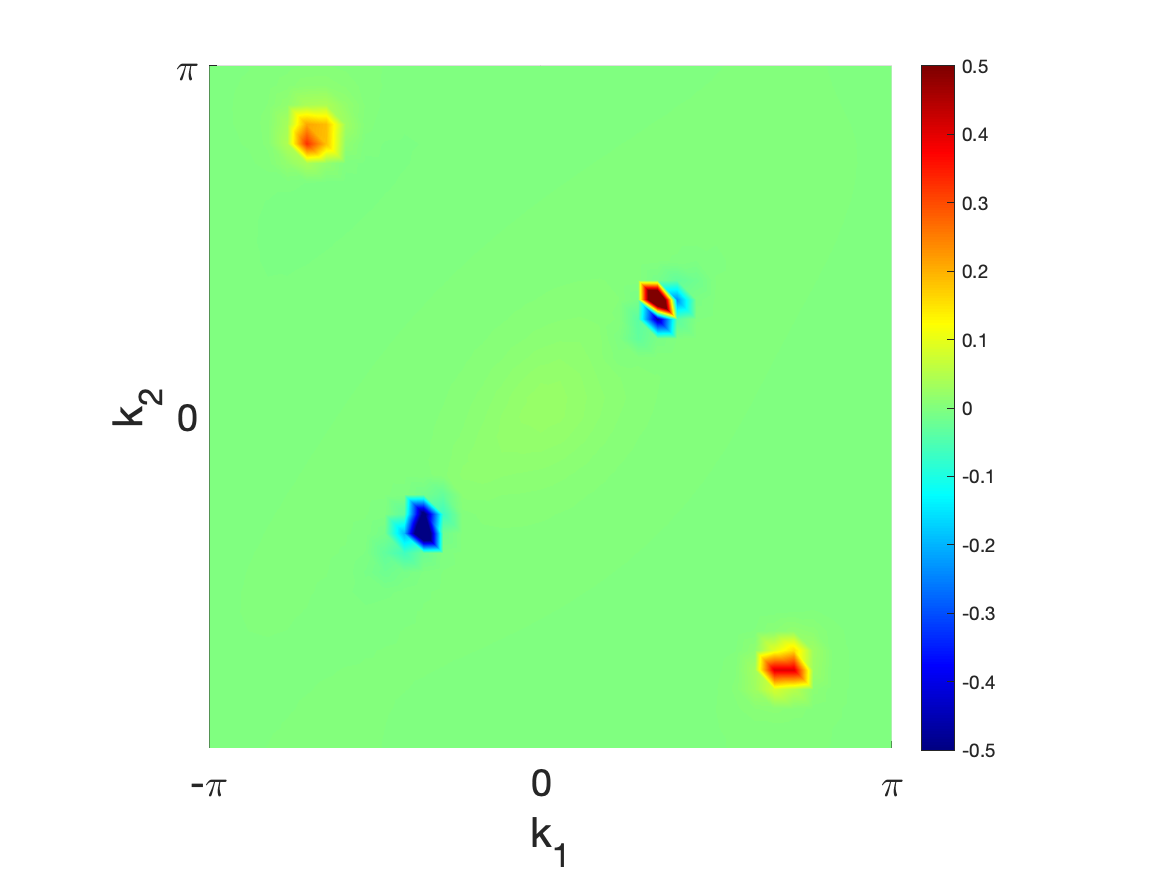}
	\caption{Non-Abelian computation of the Berry curvature for the first and second modes of the hexagonal EM duality cell described in \cite{bisharat_electromagnetic-dual_2019}, projected into the pseudospin up space of $\psi^+ = \sqrt{\epsilon_0}E_z + \sqrt{\mu_0}H_z$. The spikes at interior BZ points are numerical artifacts arising from the accidental degeneracy of the TE and TM-like mode components.}
	\label{fig:nonAbel}
\end{figure}

The results of Fig. \ref{fig:rhomSpin} show the Berry curvature calculation when one mode is projected into one of the two the pseudospin spaces, with the other mode ignored. This gives a direct look at that mode's contribution, but obscures the the effect of the other. Since they are partially degenerate, we may instead compute the non-Abelian form of the Curvature, which captures the effect of both modes, projected onto the same pseudospin space. The result of this for $\psi^+$ is shown in Fig. \ref{fig:nonAbel}. Here we see the two positive-sign peaks at both $K$ and $K'$, which, when integrated, each give $+\frac{1}{2}$. This leads directly to the expected quantum spin Hall-like physics, where the spin-Chern number is given by $C_{spin} = C_{K/K'} = +1$. The same result repeated for the $\psi^-$ space gives $C_{spin} = -1$, as expected. 

\section{Effects of Imperfect Duality and Finite Dielectric Constants}\label{app:rashba}
The ideal TB model of Eq. \ref{eq:C2KMH} assumes a perfect Kramers degeneracy across the BZ, which results in the overlapping Dirac cone-like dispersions of the edge states in the finite sample shown in Fig. \ref{fig:C2KMHribbon}. However, we note that in the electromagnetic unit cell this condition is only met with ideal electromagnetic duality, $\epsilon_r = \mu_r$. This is challenging to achieve experimentally, as there would ordinarily need to be a dielectric spacer between the two metallic sheets \cite{bisharat_electromagnetic-dual_2019}. Moreover, there is a finite capacitance of the structure that necessarily causes a slight mismatch, as seen by the bands in Fig. \ref{fig:rhomCell}(b). 

This effect does not fundamentally alter the global topology, as we can observe within the TB model. To approximate the effect of the finite dielectric in the TB model of Eq. \ref{eq:C2KMH}, we may include a Rashba-like term $H_R = \lambda_R (\sigma_x k_y - \sigma_y k_x) \ne 0$ which acts to split the Kramers degeneracy. As this acts along the off-diagonal elements of the Hamiltonian, there is will naturally impose inversion symmetry breaking, which is the case when considering the full frame + patch arrangement of the unit cell. However, we note that Rashba coupling does not inherently destroy the global topology, provided this does not close the bandgap \cite{kane_quantum_2005}.

\begin{figure}
	\centering
	\includegraphics[width=\linewidth]{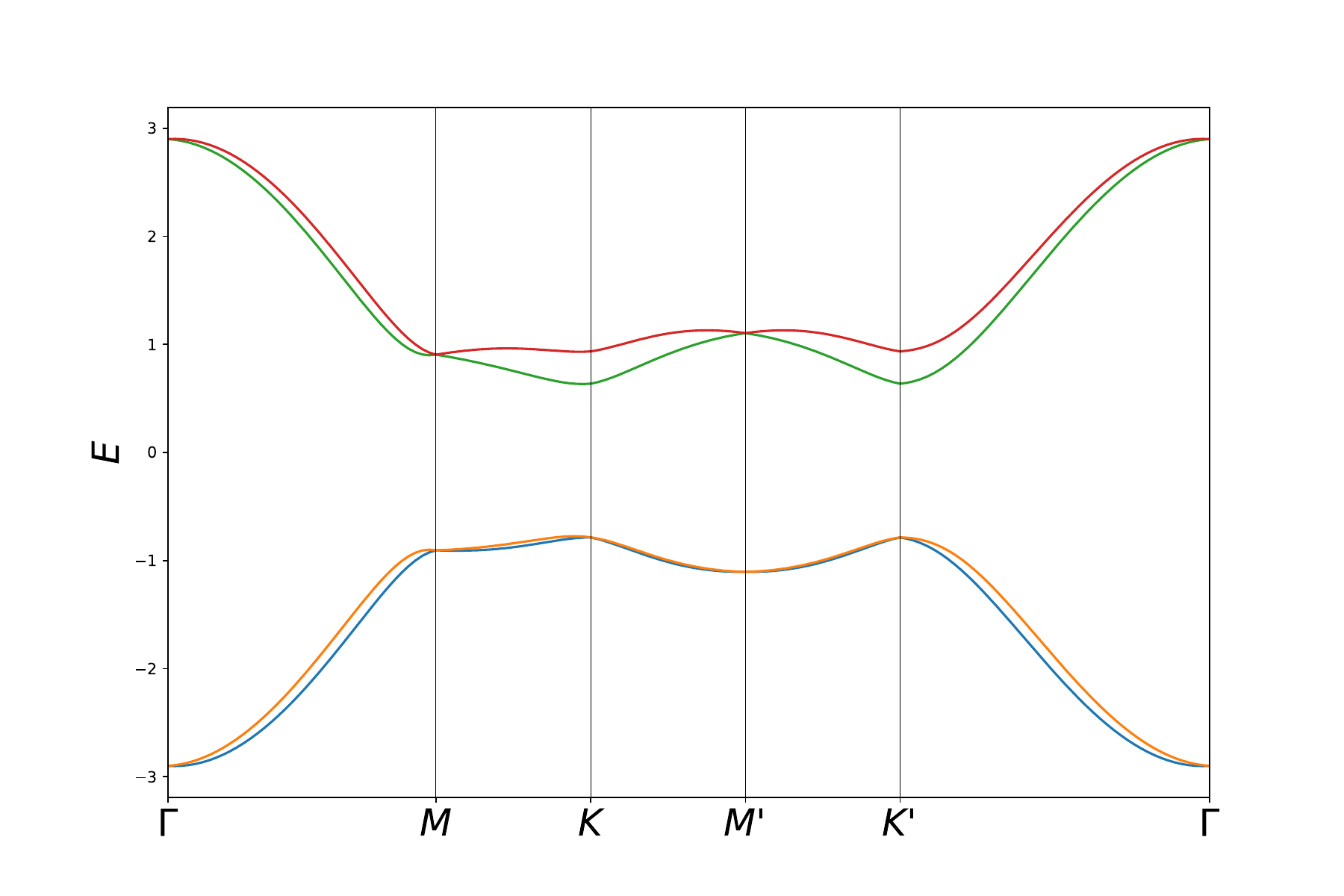}
	\caption{Effect of adding a Rashba-like spin-orbit coupling term to the TB model of Eq. \ref{eq:bloch}}
	\label{fig:rashba}
\end{figure}

An example of this for the lattice model Eq. \ref{eq:bloch}, we can add in the Rashba term to the tight-binding Hamiltonian $H(\mathbf{k},t_1^r,t_1^b,t_2) + H_R(\lambda_R)$, a representative band structure for which is given in Fig. \ref{fig:rashba}. This leads to momentum-dependent splitting of the two spin bands, but does not globally destroy the non-trivial topology \cite{kane_quantum_2005, kane_z_2005}. 

\section{Consequences of Numerical Simulation of Berry Curvature}\label{app:gauge}
The numerically simulated eigenmodes of the EM unit cell are calculated via the finite element method, which must discretize the geometry into a converged mesh. For many purposes (including band structure calculations like that shown in Fig. \ref{fig:rhomCell}(b)) the tolerance settings for the simulations is not a sensitive parameter, and most commercial solvers will generate physically meaningful results.
However, for calculations of Berry curvature and the related Wilson loop spectra presented in Fig. \ref{fig:rhomCurv} and Fig. \ref{fig:EMWilson}, small variations to the mesh can induce random phase changes between adjacent calculations that distort the distribution. As proven in \cite{fukui_chern_2005}, the four-point calculation method alleviates the requirement of maintaining a globally smooth gauge, but jumps due to this mesh-induced variation will appear like noise on the final outputs. 

As the Berry curvature corresponds to both the bandgap size and the overlap between bands, these phase distortions can lead to unintelligible results, even when using very tight tolerances. To avoid this, the phase can be smoothly maintained by forcing the solver to reuse the same mesh on all adjacent $k$ points. This manually fixes the gauge, ensuring that the topological quantities of interest smoothly vary between each plaquette, improving the visual interpretation of the results. Note that this only impacts the visuals of the plots; the real invariant quantities of the integral (for the Berry curvature) and winding (of the Wilson loop) are maintained even with a changing mesh density. 

\section{Square Unit Cells}\label{app:square}

\begin{figure}
	\centering
	\includegraphics[width=\linewidth]{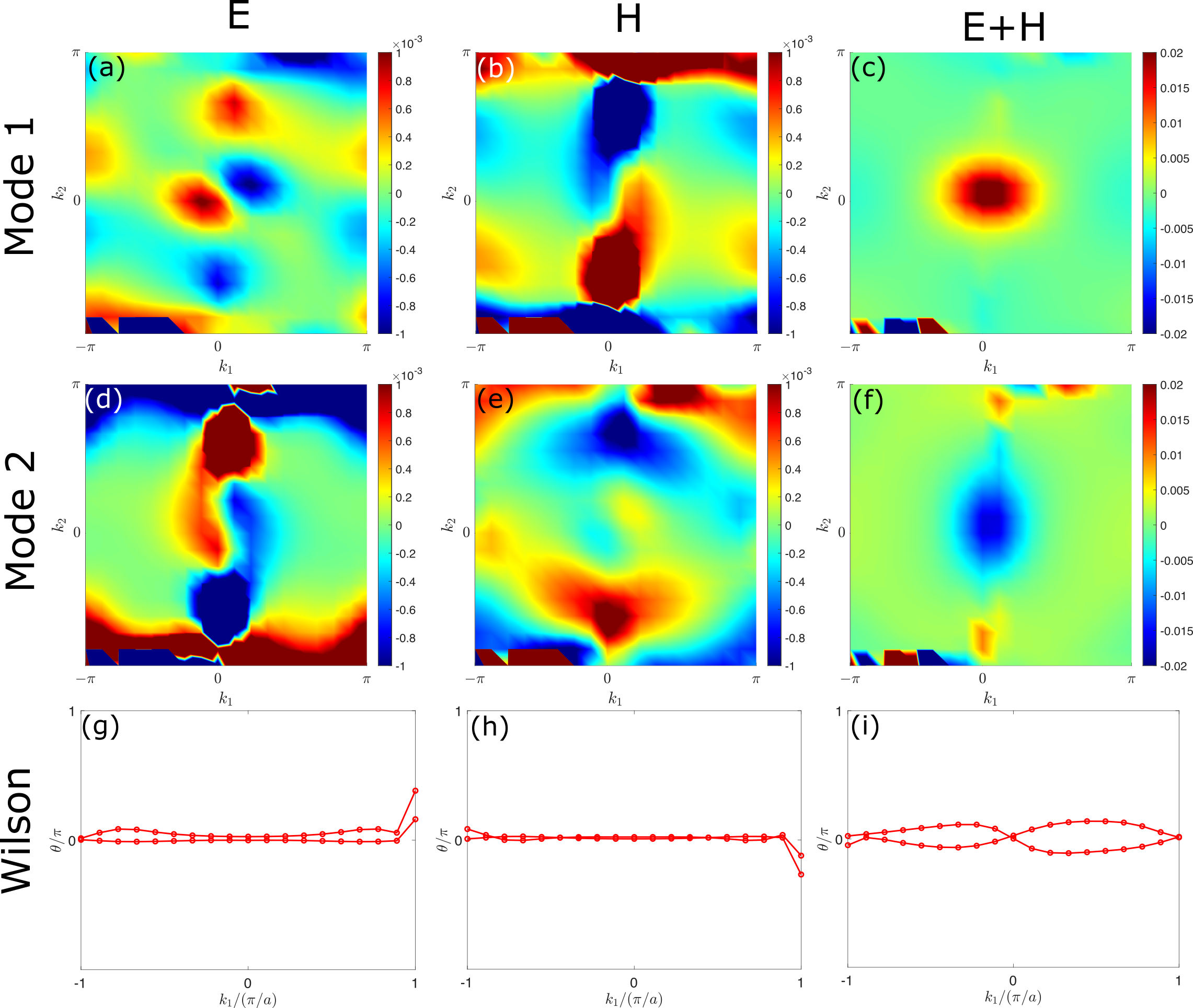}
	\caption{Topological behavior of the square EM unit cell. (a-c) shows the Berry curvature behavior of the first mode considered in isolation, using the (a) $E$ field, (b) $H$ field, and the spin-projected combination of $E+H$ fields, respectively. (d-f) show the same, but for the second mode in isolation. (g-i) show the non-Abelian Wilson loop using the $E$, $H$, and $E+H$ fields for the lowest two modes together. In all cases, the model is trivial. }
	\label{fig:square}
\end{figure}

A simulation of the duality EM unit cell structure where the angle is fixed to 90 degrees (e.g., a square) is shown in Fig. \ref{fig:square}. As explained in the main text, as the angle departs from the 60 degree point of the rhombus case, the definition of the pseudo spins starts to break down, and the topological characteristics start to degrade. In the 90 degree case shown, the pseudo spin Berry curvatures (c) and (f) show no non-trivial behavior, having only a small accumulation near the $\Gamma$ point. Likewise, we see the Wilson loop spectra collapsing down, no longer spanning the full $2\pi$ range required for non-trivial behavior. Hence, we do not expect to find non-trivial edge modes for this square cell. 

Also visible in panels (a,b,d,e) is a numerical artifact located near the $(-\pi,-\pi)$ point. As explained in Appendix \ref{app:gauge}, this arose due to a slight deviation in the mesh for those points. This is left in to illustrate the effect, as well as to indicate how the Wilson loop is a better metric to examine here: the effect of the phase discontinuity is much less, since the loop spectra is a more global quantity vs the localized behavior of the curvature. 

\bibliography{cites}

\end{document}